\documentclass[10pt]{article}
\usepackage{array}
\usepackage{rotating}
\usepackage{multirow}
\usepackage{mwe,tikz}\usepackage[percent]{overpic}
\usepackage{color}
\usepackage[colorinlistoftodos]{todonotes}
\usepackage[normalem]{ulem}
\usepackage[affil-it]{authblk}
\usepackage{amsmath}
\usepackage{fullpage}

\title{Diffractive shear interferometry for extreme ultraviolet high-resolution lensless imaging}

\author[1,2]{G. S. M. Jansen}
\author[1,2]{A. C. C. de Beurs}
\author[1,2]{X. Liu}
\author[1,2]{K. S. E. Eikema}
\author[1,2,*]{S. Witte}

\affil[1]{\small Advanced  Research  Center  for  Nanolithography  (ARCNL), Science  Park  110,  1098  XG  Amsterdam, Netherlands} \affil[2]{Dept.  of  Physics  and  Astronomy, Vrije  Universiteit, De  Boelelaan  1081,  1081  HV  Amsterdam, Netherlands\newline}

\affil[*]{email: witte@arcnl.nl}

\begin{document}
\maketitle

\begin{abstract}
We demonstrate a novel imaging approach and associated reconstruction algorithm for far-field coherent diffractive imaging, based on the measurement of a pair of laterally sheared diffraction patterns. The differential phase profile retrieved from such a measurement leads to improved reconstruction accuracy, increased robustness against noise, and faster convergence compared to traditional coherent diffractive imaging methods. We measure laterally sheared diffraction patterns using Fourier-transform spectroscopy with two phase-locked pulse pairs from a high harmonic source. Using this approach, we demonstrate spectrally resolved imaging at extreme ultraviolet wavelengths between 28 and 35~nm.
\end{abstract}



\section{Introduction}
In recent years, coherent diffractive imaging (CDI) has enabled vast progress in high-resolution microscopy \cite{chapman_coherent_2010,shapiro_biological_2005,holler_high-resolution_2017,gardner_subwavelength_2017,zurch_real-time_2014}. Contrary to traditional microscopy, CDI does not rely on lenses to form an image from scattered light emerging from a sample. Instead, CDI employs numerical phase retrieval algorithms to reconstruct an image based on the recorded diffraction pattern \cite{fienup_phase_1982,elser_phase_2003}. As the image resolution in CDI is not limited by focusing optics, it is well suited for microscopy using x-rays \cite{miao_extending_1999}, extreme ultraviolet (EUV) radiation \cite{gardner_subwavelength_2017,zurch_real-time_2014} or electrons \cite{huang_sub-angstrom-resolution_2009}. Despite the high-resolution results, the quality of the reconstructed intensity and phase of the images depends strongly on the signal-to-noise of the diffraction pattern \cite{martin_noise-robust_2012}. Furthermore, other constraints such as finite support, positivity or atomicity are often required for convergence. This has led to the development of ptychography \cite{rodenburg_phase_2004,thibault_high-resolution_2008,zhang_ptychographic_2016}, which eliminates the need for strong constraints by taking much more data in a systematic manner. 

The central challenge in CDI is to acquire knowledge of the phase of the recorded field. Performing a direct measurement of the phase is therefore beneficial, but typically does come at the cost of increased measurement complexity. The main example of such an approach is Fourier transform holography, in which the interference between a reference wave and a diffraction pattern is recorded \cite{gabor,tenner_fourier_2014,eisebitt_lensless_2004,sandberg_tabletop_2009,gauthier_single-shot_2010}. Holography allows for a simple image reconstruction which does not rely on iterative algorithms, but the image resolution and support are typically limited by the numerical aperture and wavefront of the reference wave.

Spatial phase determination of optical fields is a challenge that has been addressed in other areas as well. A specific approach that has shown promise in the context of CDI is lateral shearing interferometry (LSI) \cite{riley_laser_1977,austin_lateral_2011}, a technique that is used to reconstruct the wavefront - or phase profile - of a beam by interfering is with a sheared copy of itself. This results in an interference pattern that depends on the spatial derivative of the wavefront, which can be retrieved by spatial Fourier filtering. The wavefront can then be reconstructed by integration of the measured phase derivative. As a single LSI measurement only yields the one-dimensional derivative of the phase along the shear direction, several measurements with different shears are in principle necessary to retrieve the full 2D wavefront. Furthermore, accurate phase determination is only possible if the individual beams have smooth intensity profiles. 
The LSI phase profile can also be measured by phase shifting one of the beams and measuring the interference pattern for several phases. Isolation of the oscillating interference term then allows for direct determination of the interference phase. Such a measurement allows for measurement of much more complex interference patterns. It has been shown that a collection of shear interference patterns for varying shears allows for the full reconstruction of the original electric field \cite{falldorf_wave_2013}. Simple numerical propagation of the electric field then enables phase contrast microscopy in various geometries. 

In this article, we build upon the concept of lateral shearing interferometry to acquire differential phase information of diffraction patterns recorded with extreme ultraviolet radiation. The resulting diffraction intensity and differential phase information are then used as input for an iterative algorithm, that can reconstruct the full electric field based on a single laterally sheared diffraction pattern. The measurement and reconstruction of these diffraction patterns can be summarized as diffractive shear interferometry (DSI). In comparison with traditional coherent diffractive imaging methods, we find that our approach provides an improved reconstruction accuracy and convergence. For an experimental demonstration of the DSI approach, we measure laterally sheared diffraction patterns at several extreme ultraviolet wavelengths and numerically reconstruct high-resolution images from them. To achieve spectral resolution, we employ spatially-resolved Fourier-transform spectroscopy with a pair of phase-locked high-harmonic generation sources \cite{jansen_spatially_2016}. From these results we find that our algorithm is able to accurately reconstruct complex electric fields even in the presence of significant noise.

\section{Spatial shearing interferometry of diffraction patterns}
\subsection{Interference of diffraction patterns}
\label{dsi_math}
\begin{figure}[ht!]
\center{\includegraphics[width=0.83\columnwidth]{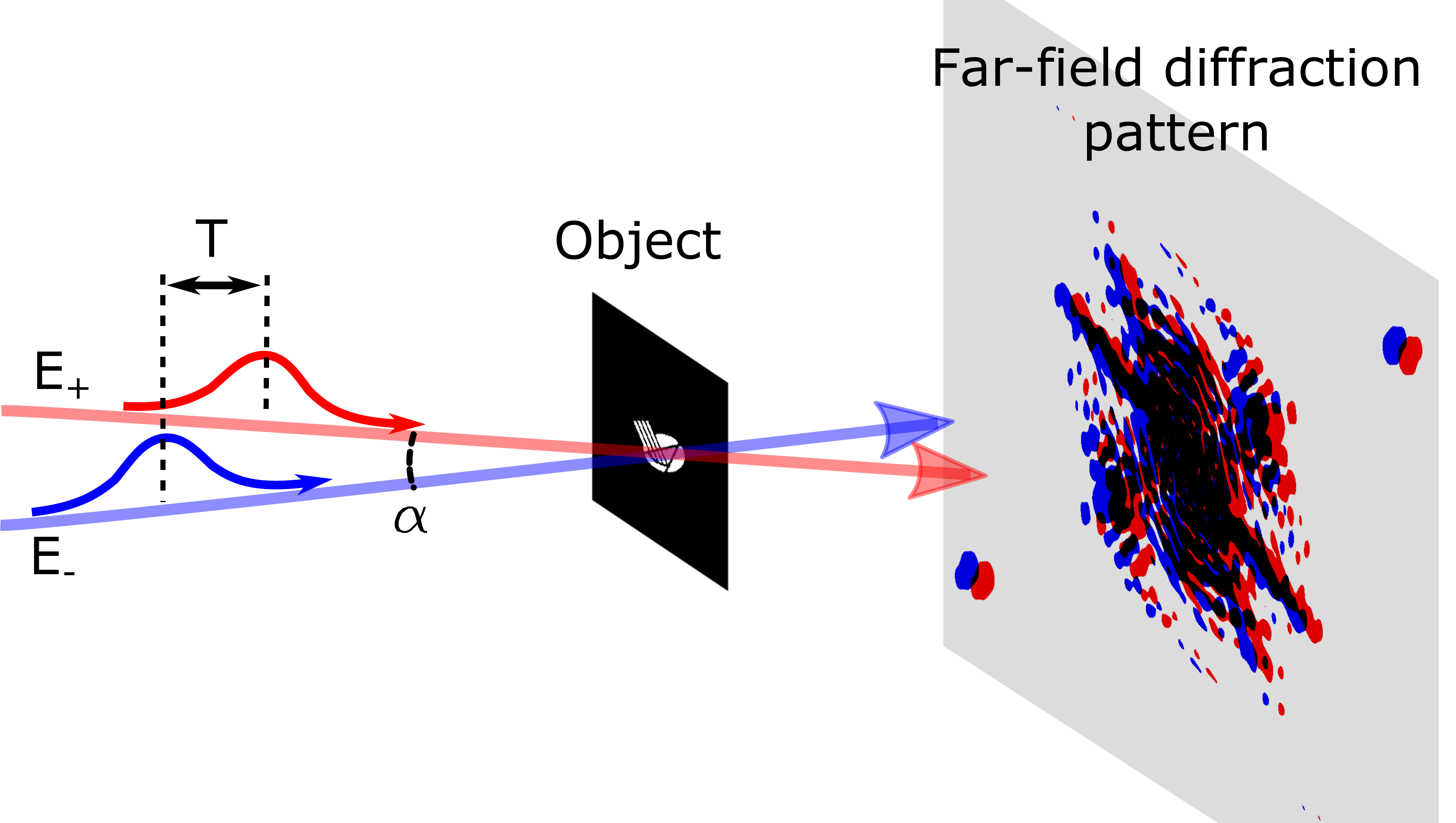}}
\label{shear_geometry}
\caption{Schematic overview of lensless imaging by diffractive shear interferometry. Two identical and coherent beams, $E_+$ and $E_-$, illuminate a transmissive object such that the angle between the beams is $\alpha$. This results in two far-field diffraction patterns on the camera which are slightly displaced relative to each other. The black region on the screen indicates where the diffraction patterns overlap and interfere.}
\end{figure}

In traditional CDI, the object is illuminated using a single beam of coherent, monochromatic light. The transmitted or reflected light scatters from the object and forms a diffraction pattern which is captured using a camera. In order to combine CDI with lateral shearing interferometry, we use two identical, mutually coherent beams to illuminate the object at slightly differing angles, as schematically depicted in Fig.~\ref{shear_geometry}. For a thin, single-scattering object, the electric fields of the beams can then be written as the electric field of the illumination multiplied by the object transfer function. If $dk$ is the wavevector corresponding to the half angle $\alpha/2$ between the two beams and $X$ the transverse position in the object plane, the electric field corresponding to a single beam appearing directly after the object can be written
\begin{equation}
E_{\pm} = \tilde{A}(x) \exp[i(\tilde{\Phi}(x) \pm dk x)] \exp(-i \omega t_{\pm}),
\label{fields_pm}
\end{equation}
where "+" and "-" correspond to the individual beams as indicated by Fig.~\ref{shear_geometry}.
Effectively, the fields in Eq.~\ref{fields_pm} consist of the amplitude $\tilde{A}(x)$ and phase $\tilde{\Phi}(x)$ of the electric field transmitted by the object, multiplied by a linear phase ramp $\exp(\pm i dk x)$ that distinguishes the two individual beams. Finally, there is a global phase term $\exp(i \omega t_{\pm})$. This electric field propagates towards the detector where the interference between $E_+$ and $E_-$
is detected. In the case of far-field diffraction, the electric field can be described by the Fourier transform of the electric field in the object plane. The detected intensity can then be written as
\begin{equation}
I(k) = A(k+dk)^2 + A(k-dk)^2 + A(k+dk)A(k-dk)\exp\{i[\Phi(k+dk)-\Phi(k-dk)+\omega T]\}+c.c.,
\label{detected_int}
\end{equation}
where the combined amplitude and phase $A(k)\exp[i\Phi(k)]$ at the camera is related by Fourier transform to the combined amplitude and phase $\tilde{A}(x)\exp[i\tilde{\Phi}(x)]$ at the object. For simplicity, the time difference between the beams is written as $T$. 
Compared to single-beam CDI, the detected intensity in DSI contains more information as it encodes the phase shear in the interference term. This will naturally lead to a more stringent camera-space constraint that can be expected to aid convergence of phase retrieval algorithms \cite{loetgering_phase_2017}. 
Yet the available information can be exploited more effectively by separating the amplitude and phase terms in the interference term $A(k+dk)A(k-dk)\exp{i[\Phi(k+dk)-\Phi(k-dk)]}$. This can be achieved by performing measurements at multiple time delays $T$, as the interference is the only term in Eq.~\ref{detected_int} that oscillates at frequency $\omega$. Going one step further, taking a series of measurements as a function of $T$ is equivalent to Fourier transform spectroscopy, and can even be used to extract interference terms for all frequencies present in the case of broadband illumination. Therefore, the proposed DSI approach is intrinsically compatible with polychromatic or broadband light sources such as high-harmonic generation, and can be used for spectrally resolved imaging at extreme ultraviolet wavelengths. 

\subsection{Image reconstruction}
\label{algorithm}
To reconstruct an image of the object, retrieval of the full electric field $A(k)\exp[i\Phi(k)]$ is required. 
Starting with the isolated interference term from Eq.~\ref{detected_int}, we will use an iterative algorithm to reconstruct the electric field. This algorithm relies on a set of  constraints to the electric field, applied in different planes connected by free-space optical propagation~\cite{chapman_coherent_2010,fienup_phase_1982,elser_phase_2003}. The first constraint is provided by the measured data: the electric field at the camera plane should be consistent with the measured result. 
For our second constraint we will use a finite support; the electric field in the object plane is only non-zero in a certain limited window. 

In traditional CDI, the most used camera plane operator is the modulus constraint: the amplitudes of the electric-field estimate are set to the measured values, while the estimated phases are preserved. However, in DSI the modulus constraint is not the most suitable operator for reconstructing interferometrically sheared diffraction patterns. This is because the measured intensity pattern  $A(k+dk)A(k-dk)$ is not equal to the intensity $A(k)^2$ of the electric field to be reconstructed. Even though a modulus constraint based on Eq.~\ref{detected_int} may be envisaged, it does not take into account the available phase information in an optimal way. 
We therefore derive a new camera plane operator that makes more efficient use of the amplitude and phase information available in DSI. 

Starting with the $n^{th}$ guess of the electric field
\begin{equation}
E_n(k) = A_n (k) \exp[i\Phi_n (k)]
\end{equation}
at the camera and the complex measured interference pattern
\begin{equation}
M(k) = A(k+dk)A(k-dk)\exp\{i(\Phi(k+dk)-\Phi(k-dk)]\},
\label{measured_field}
\end{equation} 
it can be seen that division of the measured data by a sheared copy of the electric field guess yields a new electric field guess
\begin{equation}
E_{n+1}(k + dk) = \frac{M(k)}{E_{n} ^* (k - dk)}.
\end{equation}
If the original guess of the electric field is accurate, the new electric field guess is in fact equal to a shifted version of the electric field. Therefore, division of the measured data by a negatively sheared electric field guess yields a positively sheared electric field guess. This operation forms the basis of a useful camera-space constraint for interferometrically sheared diffraction pattern reconstruction. A general camera-space constraint can be written as
\begin{equation}
E_{n+1} (k) = (1-\beta) E_n (k) + \frac{\beta}{2}\left[\frac{M(k-dk)E_n (k - 2dk)}{|E_n(k - 2dk)|^2+\alpha^2}+\frac{M^*(k+dk)E_n (k + 2dk)}{|E_n(k + 2dk)|^2+\alpha^2}\right],
\end{equation}
which is a linear combination of the old guess and the average of the new guesses for the positively and negatively sheared electric fields. The numerical constant $\beta$ determines the strength of the correction to the electric field guess and is typically set to 0.9. Instead of a direct division by the electric field $E$, we multiply by $E^*/(|E|^2+\alpha^2)$, where $\alpha$ is a regularization constant that prevents errors arising from division by zero. 

In combination with a finite object support, the presented camera-space constraint is sufficient for retrieval of the full electric field. A basic approach to include a support constraint in the algorithm is using the error-reduction method, where all values outside of the support are set to zero. The output-output algorithm and hybrid input-output algorithms provide two alternatives which have been shown to provide different convergence properties \cite{fienup_phase_1982,marchesini_invited_2007}. 

\subsection{Comparison between DSI and single-beam CDI}
\begin{figure}[t!]
\centering\includegraphics[width=\linewidth]{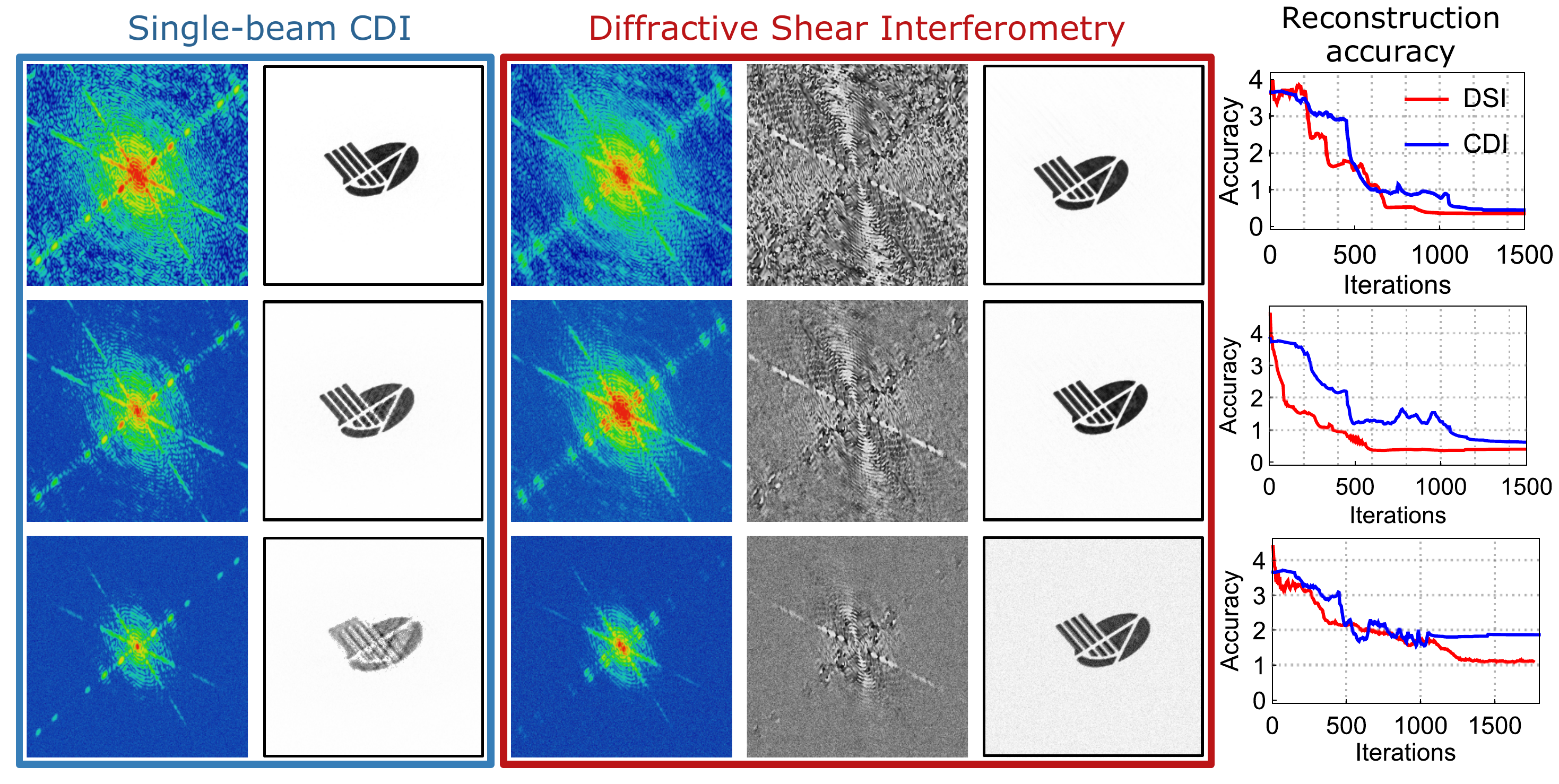}
\caption{Comparison of traditional single-beam CDI with the proposed diffractive shear interferometry. Each row compares the two methods for identical signal-to-noise levels. The first two columns show the simulated diffraction patterns and reconstructions, respectively, for single-beam CDI. Columns three and four show the simulated DSI amplitude and phase. Column five shows the image retrieved using our algorithm. Finally, the last column compares the accuracy of both methods. The accuracy is calculated from the mean absolute difference between reconstruction and original image, averaged over the number of pixels. For clarity, the accuracy calculation only considers the direct vicinity of the object.}
\label{cdi_vs_dsi}
\end{figure}

To investigate the efficiency of the proposed phase-retrieval algorithm, we performed simulations comparing DSI to traditional single-beam CDI methods. Example datasets for CDI are produced by simulating far-field diffraction patterns with Poisson noise. We simulated several diffraction patterns with varying signal-to-noise ratio (SNR) by adjusting the total number of collected photons. To account for camera readout noise, we added Gaussian background noise with a standard deviation of 10 counts. Starting with a wide initial support, we reconstructed the image using both hybrid input-output and error-reduction in an alternating fashion. The shrinkwrap procedure was used to adaptively update the support \cite{marchesini_xray_2003}. The results of these simulations can be seen the first two columns of Fig.~\ref{cdi_vs_dsi}. In this figure, each row has a different SNR. The top row is simulated such that we get $10^8$ counts in the brightest pixel, leading to a SNR of $10^4$ in the center of the diffraction pattern. The second and third row have SNRs of $1.8 \cdot 10^3$ and $0.6 \cdot 10^3$ respectively. 

For DSI, we simulated several datasets with noise levels similar to the CDI simulations. This was achieved by adding Poisson noise and a Gaussian background to the shear interferometry signal. To also account for noise in the phase of the simulated interference pattern, we added a random phase factor to the total of Poisson and Gaussian noise. The amplitude and phase of the simulated diffraction patterns can be seen in the third and fourth columns of Fig.~\ref{cdi_vs_dsi} respectively. We verified that this method leads to realistic interference signals by comparing them the result of a full simulated Fourier transform scan. Furthermore, we compared the simulated data to real measured EUV shearing interferometry data (section~\ref{actual_measurement}). 

Reconstructions of the simulated DSI signals are shown in the fifth column of Fig.~\ref{cdi_vs_dsi}. These reconstructions were obtained by combining the algorithm presented in section~\ref{algorithm} with a shrinkwrap procedure to find the support of the image. In order to compare these results to those obtained from the CDI simulations, we calculate the accuracy of the reconstructions. This is defined as the RMS difference between the reconstruction and input image, averaged over the number of pixels  and only calculated for the pixels in the direct vicinity of the original object. This calculation is corrected for spatial shifts and a global phase offset, which are free parameters for CDI. For DSI, we find that there is a limited number of spatial shifts and phase offsets compatible with the measured data. This is directly related to the phase information in the recorded diffraction pattern, as a shift of the object translates to a phase tilt in the far-field diffraction pattern. After shearing according to formula~\ref{measured_field}, this phase tilt reduces to a phase offset in the measured data. Reversing the process shows that a phase offset in the measured data leads to a shift of the reconstructed object, and any image reconstruction has to match this constraint.

From the reconstruction accuracy comparison in the sixth column of Fig.~\ref{cdi_vs_dsi}, it is clear that DSI consistently leads to a better solution than single-beam CDI. In addition, the reconstruction of DSI patterns appears to converge slightly faster than the reconstruction of single-beam CDI patterns. 
There are various parameters which influence the performance of DSI reconstruction, of which the value of the shear is the most critical. For the simulations presented in Fig.~\ref{cdi_vs_dsi}, we assumed that the shear was known accurately. For real measurements, accurate initial knowledge of the shear may not be possible, especially in cases where the shear has to be known with sub-pixel accuracy. In such cases, it is possible to extend the phase retrieval algorithm with a shear optimization step. As will be mentioned in section~\ref{actual_measurement}, for our present reconstructions we have used a manual search to find the correct shear.

Furthermore, the value of the shear has a strong influence on the measured signal and therefore on the retrieval process. If the shear is reduced to zero, the shear interferometry signal reduces to the single-beam far-field diffraction pattern and the phase information is reduced to zero. As the shear is increased, both the intensity of the diffraction pattern and the phase information become more complex-structured. Finally, very large shears lead to a reduced overlap between the diffraction patterns, leading to a weaker signal that is more sensitive to noise. We simulated several measurements with different shears, and found that larger shears lead to slightly better results and faster convergence, provided that the SNR remained sufficiently high. The shear used in Fig.~\ref{cdi_vs_dsi} was approximately equal to one speckle of the diffraction pattern (the inverse of the object size), which is found to be a good compromise between signal strength and noise sensitivity.

\section{Experimental demonstration of diffractive shear interferometry using high-harmonics}
\label{actual_measurement}

\begin{figure}[ht!]
\centering\includegraphics[width=0.8\linewidth]{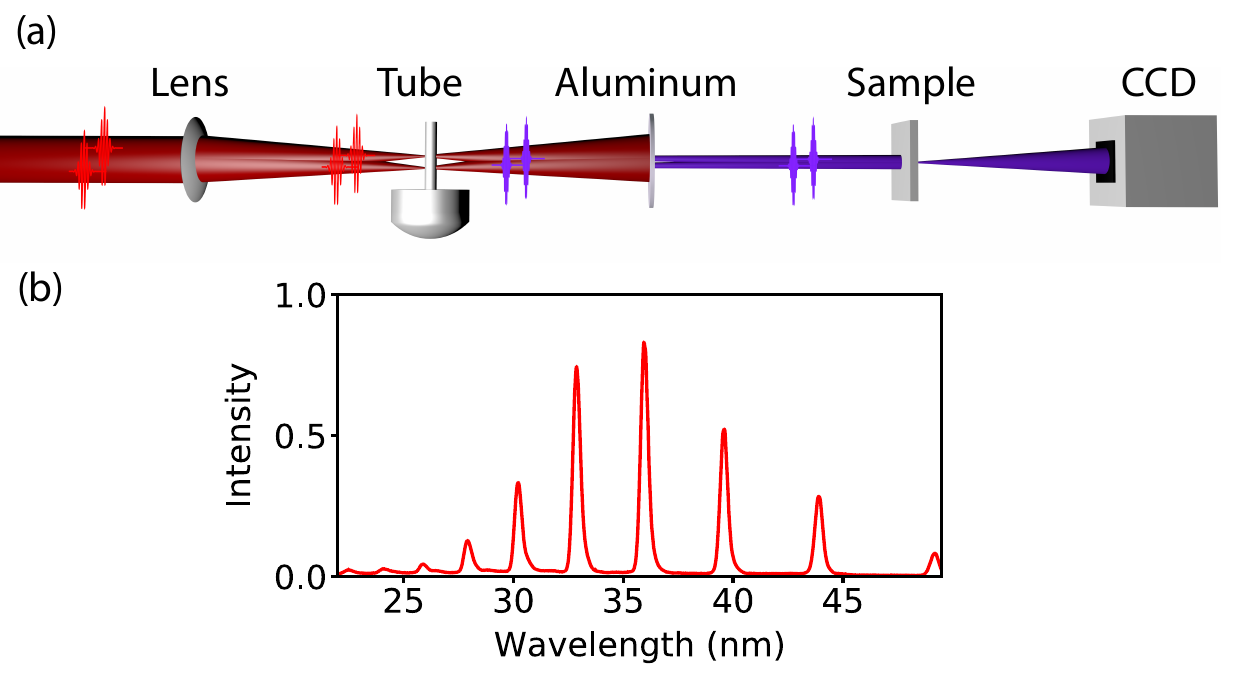}
\caption{(a) Schematic overview of the setup used for EUV Fourier-transform interferometry. The output of a common-path interferometer is focused by the lens into a gas jet confined to tube. The resulting EUV pulse pair is separated from the infrared using an Aluminum filter and detected using an Andor Ikon-L CCD camera. A transmissive object can be positioned between the Aluminum filter and the camera using a remotely controlled translation stage. (b) Typical high-harmonic spectrum generated in Argon, measured through Fourier-transform spectroscopy in the same setup without an object in the beam.}
\label{setup}
\end{figure}

As already noted in section~\ref{dsi_math}, a promising approach to measuring DSI signals is through the use of a setup for Fourier-transform spectroscopy. Such a measurement enables coherent diffractive imaging using all wavelengths present in the illumination. We implement this scheme by employing a phase-locked pair of high-harmonic generation (HHG) sources to perform wavelength-resolved microscopy at EUV wavelengths. The phase-locked EUV pulse pair is produced by HHG upconversion of tightly phase-locked pairs of infrared driving pulses that have been produced by an ultrastable common-path interferometer~\cite{jansen_spatially_2016}. Typical parameters of these infrared pulses are a central wavelength of 800~nm, a pulse energy of 1~mJ in each of the beams, a 300~Hz repetition rate and a pulse duration of 25~femtoseconds. A basic layout of the setup used for HHG and subsequent DSI imaging of samples in a transmission geometry is presented in Fig.~\ref{setup}(a). A typical EUV spectrum generated in Argon measured through Fourier-transform spectroscopy is also shown in Fig.~\ref{setup}(b). It is important to note that the HHG spectrum can change significantly due to small changes in the driving laser alignment. 

\subsection{Fourier-transform holography}
\begin{figure}[ht!]
\centering\includegraphics[width=0.86\linewidth]{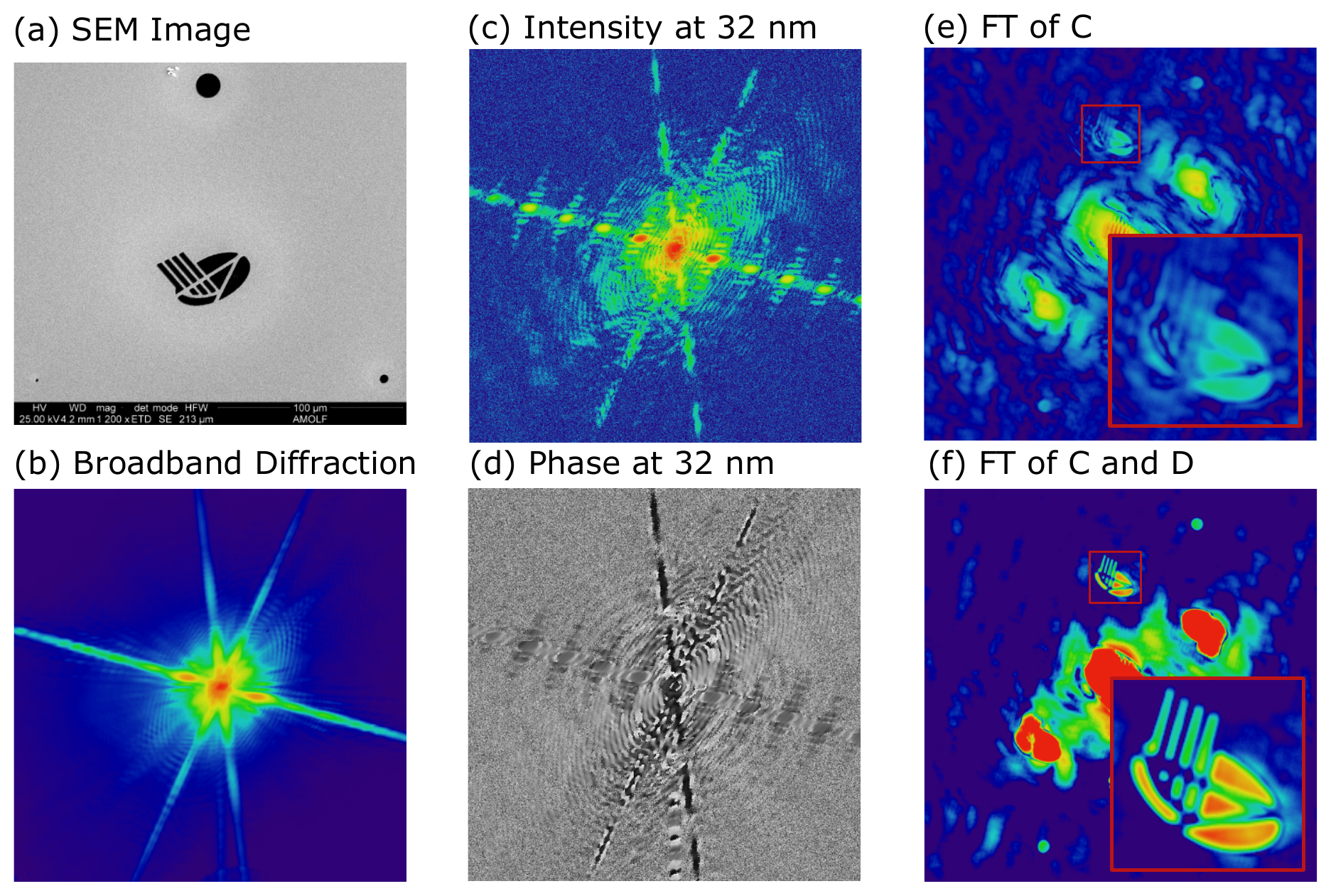}
\caption{Extreme ultraviolet DSI of a transmissive sample with multiple holographic references.(a) Scanning electron microscope image of the sample used for the initial measurements at EUV wavelengths. The sample consists of our institute logo and three circular apertures with diameters 12, 4 and 1~$\mu$m respectively acting as holographic references. (b) Broadband EUV transmission of the sample. (c) Intensity at 32~nm retrieved from the FTS-scan. (d) Phase of the signal at 32~nm. (e) Hologram acquired using just the intensity at 32~nm as shown in (c). (f) Hologram acquired using both intensity and phase information at 32~nm. The inset show the hologram arising from interference with the 4~$\mu$m aperture.}
\label{holography}
\end{figure}

As a first experiment, we used focused ion-beam milling to fabricate the sample shown in Fig.~\ref{holography}(a). The sample consists of our institute logo and in addition three circular apertures with diameters 12, 4 and 1~$\mu$m, respectively. The apertures act as references for Fourier-transform holography (FTH) by providing a spherical wave which interferes with the diffraction pattern arising from the logo \cite{eisebitt_lensless_2004}. For such a sample, a spatial Fourier-transform of the far-field diffraction pattern directly yields multiple images of the sample. The resolution of these images is determined by the diameter of the associated reference aperture. Therefore, FTH can provide an initial low-resolution guess of the image from which it is possible to determine the support. It is then possible to use phase-retrieval methods to improve the image resolution and contrast \cite{tenner_fourier_2014,capotondi_scheme_2012}. 

The results obtained with only FTH
are shown in Fig.~\ref{holography}. We used Fourier-transform spectroscopy to retrieve monochromatic shear interferometry signals for several individual high-harmonics. The spectral resolution in this experiment is 96~THz, providing an effective bandwidth $\Delta \lambda / \lambda$   of approximately $1/100$ for the retrieved shear interferometry signals around 30~nm wavelength. As an example, we show the amplitude and phase of the shear interferometry signal for the 25th harmonic at 32~nm in Fig.~\ref{holography}(c) and (d) respectively. To illustrate the importance of the phase pattern, Fig.~\ref{holography}(e) shows the hologram calculated from the amplitude data alone, which has low contrast and contains clear distortions. We find that it is possible to retrieve a good-quality hologram from the shear interferometry signal if both measured amplitude and phase are used to calculate the hologram, as shown in Fig.~\ref{holography}(f). Note that the holograms arising from the 1~$\mu $m aperture are not visible, which is probably due to limited signal-to-noise.

\subsection{DSI reconstruction}
\begin{figure}[ht!]
\centering\includegraphics[width=\linewidth]{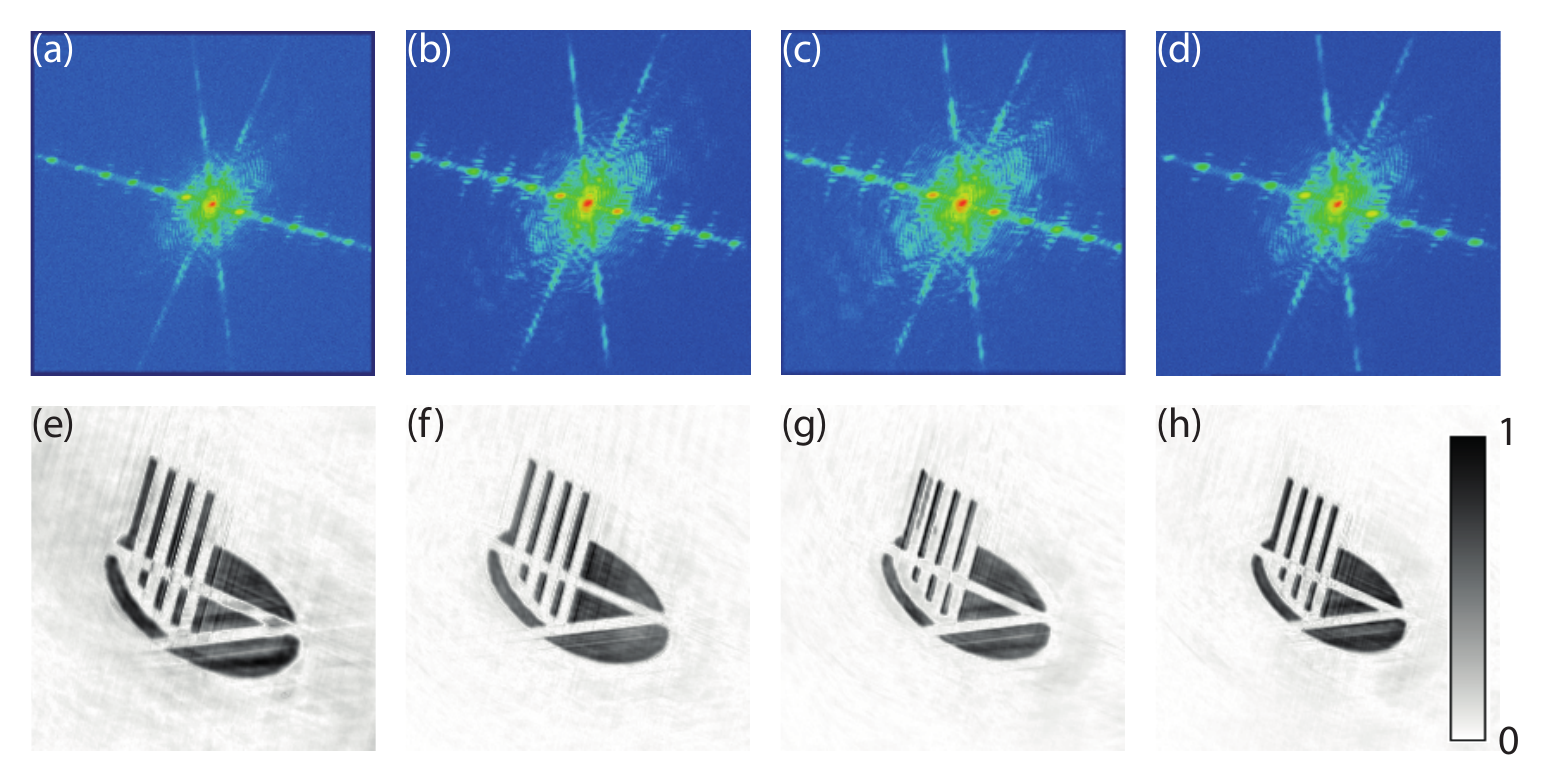}
\caption{DSI reconstruction results of the multi-wavelength data presented in Fig.~\ref{holography}. (a-d) for the ARCNL research center logo. Measured amplitude of the DSI patterns for the 29th, 27th, 25th and 23rd harmonics with wavelengths 28~nm, 30~nm, 32~nm and 35~nm respectively, shown on a logarithmic color scale. (e-h) Reconstructed images for the 29th to 23rd harmonics respectively, show on a linear grayscale. The images all have a height and width of 200 pixels, while the sample has a width of 40 ~$\mu$m. The difference in magnification follows directly from the differences in wavelength between the images.}
\label{arcnl_logos}
\end{figure}

The holography results provide a decent starting point for electric field reconstruction using DSI. For this reconstruction the holography result is used to determine an initial support. The determination of the initial support is slightly complicated by the presence of the reference apertures, as it is important to position the supports for the apertures at the correct positions relative to each other and to the main object. To accommodate for slight errors in this process, we start the image reconstruction with an object support that is larger than the image obtained through holography. The object support is determined from the hologram by thresholding and expanding the result by a few pixels. We found that applying the object support using a linear combination of error reduction and hybrid input-output leads to the best convergence. 

As with the DSI simulations, we combine the image reconstruction with a shrinkwrap routine. In addition, we performed a manual parameter search to find the optimal value for the shear. A simple search algorithm was used to find the phase offset in the measured data. As shown in Fig.~\ref{arcnl_logos}, we were able to reconstruct high-quality images using this approach for four individual high-harmonics between 28 and 35~nm from a single measurement. 

\subsection{DSI imaging of complex objects}
\label{griffin_section}
\begin{figure}[ht]
\centering\includegraphics[width=\linewidth]{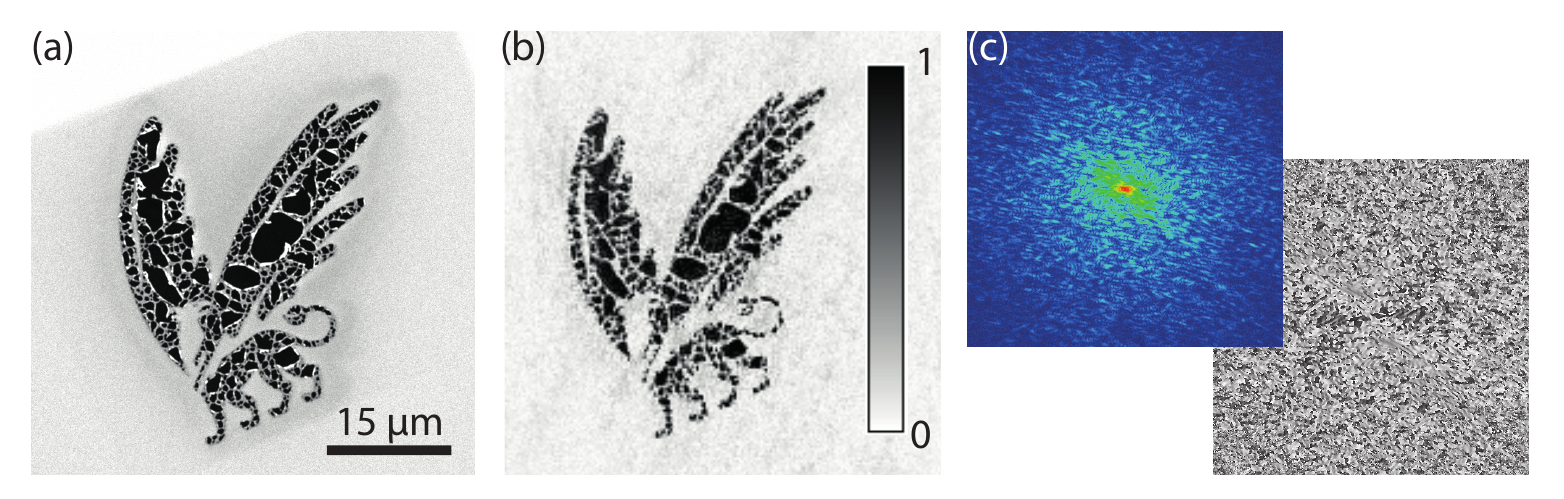}
\caption{Results obtained for a complex `griffin' sample without holographic references. (a) SEM image of the sample as described in section~\ref{griffin_section}. (b) DSI reconstruction of the sample at 34~nm wavelength, shown on a linear colorscale. (c) Measured amplitude (logarithmic false color) and phase (gray) of the DSI pattern used to obtain (b).
}
\label{griffin}
\end{figure}

To verify that DSI also works for more complicated physical data, we fabricated another sample, shown in Fig.~\ref{griffin}(a). It was produced by focused ion-beam (FIB) milling in a 100~nm gold layer on top of a freestanding 15~nm silicon nitride film. The settings of the FIB caused it to strip most of the gold while leaving a fine and irregular silicon nitride mesh. As the mesh is partially transparent to EUV radiation, this yields a sample that has a complicated pattern of transmission levels. Furthermore, the sample did not contain holographic reference apertures. 

Similar to the measurement of the holography sample, we are able to obtain monochromatic interference patterns for several wavelengths through an FTS scan. As an example, Fig.~\ref{griffin}(c) shows the amplitude and phase of the DSI pattern obtained for the 25th harmonic at 34~nm. In this case, the spectral resolution was 81~THz, yielding an effective bandwidth of $1/109$ at 34~nm wavelength.
For the reconstruction, the initial object support was now retrieved from the autocorrelation of the sample. The rest of the image reconstruction process was similar to that of the ARCNL logo sample, using a combination of error reduction and hybrid input-output in the object space. The final result is shown in Fig.~\ref{griffin}(b). By calculating the phase-retrieval transfer function, we find that the final result has a resolution of approximately 0.27~$\mu$m.
This resolution corresponds to the diffraction limit of the captured data. Comparing the SEM and DSI images, it is clear that DSI is able to reconstruct the full complexity of the sample including the partial transmission of the silicon nitride mesh. This demonstrates that DSI is a promising technique to image complex isolated samples.

\section{Conclusions}

In conclusion, we have developed the method of diffractive shear imaging, in which the full electric field of a diffraction pattern is reconstructed based on the measurement of a single sheared diffraction pattern. Comparing the algorithm to standard phase retrieval methods for traditional CDI, we find that our method consistently yields more accurate results. In addition, the phase retrieval process converges slightly faster than traditional approaches. As DSI signals can be easily measured using spatially-resolved Fourier-transform spectroscopy, this approach is ideally suited for multi-wavelength coherent diffractive imaging~\cite{Witte_2014}. In particular, this approach is interesting for CDI using high-harmonic generation sources, which produce a broad range of narrowband harmonics at extreme ultraviolet wavelengths. We have demonstrated high-resolution microscopy on two different samples using several high-harmonics at wavelengths between 28 and 35~nm.
There are several possible extensions which may lead to an even more versatile imaging technique. For example, rotation of the sample enables the measurement of shear interferometry signals at different effective shears. A set of these measurements can greatly enhance the image retrieval, and reduce the need for a well-defined object support. Furthermore, combining diffractive shear interferometry with ptychographic techniques can lead to a greater field of view while still preserving spectral sensitivity.

\section*{Funding}
The project has received funding from the European Research Council (ERC) (ERC-StG 637476) and the Netherlands Organisation for Scientific Research (NWO). 

\section*{Acknowledgments}
We thank D. E. Boonzajer Flaes for useful discussions and R. Kortekaas and N. Noest for technical assistence.


\begin{thebibliography}{1}
\newcommand{\enquote}[1]{``#1''}
\bibitem{chapman_coherent_2010}
H.~N. Chapman and K.~A. Nugent, \enquote{Coherent lensless X-ray imaging,}  Nature Photonics \textbf{4}, 844-839 (2010)
\bibitem{shapiro_biological_2005}
D. Shapiro, P. Thibault, T. Beetz, V. Elser, M. Howells, C. Jacobsen, J. Kirz, E. Lima, H. Miao, A.~M. Neiman and David Sayre, \enquote{Biological imaging by soft x-ray diffraction microscopy,} PNAS \textbf{102}(43), 15343-15346 (2005)
\bibitem{holler_high-resolution_2017}
M.~Holler, M.~Guizar-Sicairos, E.~H.~R.~Tsai, R.~Dinapoli, E.~M\"uller, O.~Bunk, J.~Raabe and G.~Aeppli, \enquote{High-resolution non-destructive three-dimensional imaging of integrated circuits,} Nature \textbf{543}, 402-406 (2017)
\bibitem{gardner_subwavelength_2017} 
D.~F. Gardner, M.~Tanksalvala, E.~R. Shanblatt, X.~Zhang, B.~R. Galloway, C.~L.  Porter, R.~Karl~Jr, C.~Bevis, D.~E. Adams, H.~C. Kapteyn, M.~M. Murnane and  G.~F. Mancini, \enquote{Subwavelength coherent imaging of periodic samples using a 13.5 nm tabletop high-harmonic light source,} Nature Photonics \textbf{11}, 259-263 (2017)
\bibitem{zurch_real-time_2014}
M.~Z\"urch, J.~Rothhardt, S.~H\"adrich, S.~Demmler, M.~Krebs, J.~Limpert, A.~T\"unnermann, A.~Guggenmos, U.~Kleineberg and C.~Spielmann, \enquote{Real-time and {Sub}-wavelength {Ultrafast} {Coherent} {Diffraction}  {Imaging} in the {Extreme} {Ultraviolet},} Scientific Reports \textbf{4}, 7356 (2014).
\bibitem{fienup_phase_1982}
 J.~R. Fienup, \enquote{Phase retrieval algorithms: a comparison,} Appl. Opt. \textbf{21}(15), 2759-2769 (1982).
\bibitem{elser_phase_2003}
V.~Elser, \enquote{Phase retrieval by iterated projections,} J. Opt. Soc. Am. A \textbf{20}(1), 40-55 (2003).
\bibitem{miao_extending_1999}
J.~Miao, P.~Charalambous, J.~Kirz, and D.~Sayre, \enquote{Extending the  methodology of {X}-ray crystallography to allow imaging of micrometre-sized non-crystalline specimens,} Nature \textbf{400}, 342-344 (1999).
\bibitem{huang_sub-angstrom-resolution_2009}
W.~J. Huang, J.~M. Zuo, B.~Jiang, K.~W. Kwon, and M.~Shim,  \enquote{Sub-\aa ngstr\"om-resolution diffractive imaging of single nanocrystals,} Nature Physics \textbf{5}, 129-133 (2009).
\bibitem{martin_noise-robust_2012}
A.~V. Martin \textit{et al.},\enquote{Noise-robust coherent diffractive imaging with a single  diffraction pattern,} Opt. Express \textbf{20}(15), 16650-16661 (2012).
\bibitem{rodenburg_phase_2004}
J.~M. Rodenburg and H.~M.~L. Faulkner, \enquote{A phase retrieval algorithm for shifting illumination,} Appl. Phys. Lett. \textbf{85}(20), 4795-4797 (2004)
\bibitem{thibault_high-resolution_2008}
P. Thibault, M. Dierolf, A. Menzel, O. Bunk, C. David, F. Pfeiffer, \enquote{High-Resolution Scanning X-ray Diffraction Microscopy,} Science \textbf{321}(5887), 379-382 (2008)
\bibitem{zhang_ptychographic_2016}
B.~Zhang, D.~F.~Gardner, M.~H.~Seaberg, E.~R.~Shanblatt, C.~L.~Porter, R.~Karl, C.~A.~Mancuso, H.~C.~Kapteyn, M.~M.~Murnane, and D.~E.~Adams , \enquote{Ptychographic hyperspectral spectromicroscopy with an extreme ultraviolet high harmonic comb,} Opt. Express \textbf{24}(16), 18745 (2016)
\bibitem{gabor}
D. Gabor, "A new microscopic principle," Nature 161, 777-778 (1948)
\bibitem{tenner_fourier_2014}
V.~T. Tenner, K.~S.~E. Eikema and S. Witte, "Fourier transform holography with extended references using a coherent ultra-broadband light source," Opt. Express \textbf{22}, 25397-25409 (2014)
\bibitem{eisebitt_lensless_2004}
S. Eisebitt, J. L\"uning, W.~F. Schlotter, M. L\"orgen, O. Hellwig, W. Eberhardt and J. St\"ohr, "Lensless imaging of magnetic nanostructures by X-ray spectro-holography," Nature \textbf{432}, 885-888 (2004)
\bibitem{sandberg_tabletop_2009}
R.~L. Sandberg, D.~A. Raymondson, C. La-O-Vorakiat, A. Paul, K.~S. Raines, J. Miao, M.~M. Murnane, H.~C. Kapteyn, and W.~F. Schlotter, \enquote{Tabletop soft-x-ray Fourier transform holography with 50~nm resolution,} Opt. Lett. \textbf{34}(11) 1618-1620 (2009)
\bibitem{gauthier_single-shot_2010}
D. Gauthier, M. Guizar-Sicairos, X. Ge, W. Boutu, B. Carr\'e, J.~R. Fienup, and H. Merdji, \enquote{Single-shot Femtosecond X-Ray Holography Using Extended References,} Phys. Rev. Lett. \textbf{105}, 093901 (2010)
\bibitem{riley_laser_1977}
M.~E. Riley and M.~A. Gusinow, "Laser beam divergence utilizing a lateral shearing interferometer," Appl. Opt. \textbf{16}(10), 2753-2756 (1977)
\bibitem{austin_lateral_2011}
D.~R. Austin, T. Witting, C.~A. Arrell, F. Frank, A.~S. Wyatt, J.~P. Marangos, J.~W.~G. Tisch, and I.~A. Walmsley, "Lateral shearing interferometry of high-harmonic wavefronts," Opt. Lett. \textbf{36}(10), 1746-1748 (2011)
\bibitem{loetgering_phase_2017}
L.~Loetgering, H. Froese, T. Wilhein, and M. Rose, "Phase retrieval via propagation-based interferometry," Phys. Rev. A \textbf{95}, 033819 (2017)
\bibitem{falldorf_wave_2013}
C. Falldorf, C. von Kopylow and R.~B. Bergmann, "Wave field sensing by means of computational shear interferometry," J. Opt. Soc. Am. A \textbf{30}(10) 1905-1912 (2013)
\bibitem{jansen_spatially_2016}
G.~S.~M. Jansen, D. Rudolf, L. Freisem, K.~S.~E. Eikema and S. Witte, \enquote{Spatially resolved Fourier transform spectroscopy in the extreme ultraviolet,} Optica \textbf{3}(10), 1122-1125 (2016)
\bibitem{marchesini_invited_2007}
S.~Marchesini, \enquote{Invited Article: A unified evaluation of iterative projection algorithms for phase retrieval,} Rev. Sci. Instrum. \textbf{78}, 011301 (2007)
\bibitem{capotondi_scheme_2012}
F. Capotondi, E. Pedersoli, M. Kiskinova, A.~V. Martin, M. Barthelmess and H.~N. Chapman, \enquote{A scheme for lensless X-ray microscopy combining coherent diffraction imaging and differential corner holography,} Opt. Express \textbf{20}(22), 25152-25160 (2012)
\bibitem{marchesini_xray_2003}
S. Marchesini, H. He, H.~N. Chapman, S.~P. Hau-Riege, A. Noy, M.~R. Howells, U. Weierstall, and J.~C.~H. Spence,"X-ray image reconstruction from a diffraction pattern alone," Phys. Rev. B \textbf{68} 130101(R) (2003)
%
\bibitem{Witte_2014}
S. Witte, V.~T. Tenner, D.~W.~E. Noom, K.~S.~E. Eikema, "Lensless diffractive imaging with ultra-broadband table-top sources: from infrared to extreme-ultraviolet wavelengths," Light Sci. Appl.\textbf{3}, e163 (2014)


\end{thebibliography}
\end{document}